# Spectrometry of the Urban Lightscape

Christopher Small
*Lamont Doherty Earth Observatory*
*Columbia University*
*Palisades, NY 10964  USA*

*csmall@columbia.edu*

NASA's Gateway to Astronaut Photography of Earth contains over 30000 photos of ~2500 cataloged urban lightscapes (urban night lights) taken from the International Space Station. Over 100 of these multispectral DSLR photos are of sufficient spatial resolution, sharpness and exposure to be used for broadband spectral characterization of urban lightscapes. Analysis of simulated atmospheric transmissivity from the MODTRAN radiative transfer model shows that spectral slopes of transmissivity spectra are relatively insensitive to choice of model atmosphere, with variations in atmospheric path length and aerosol optical depth primarily affecting the bias of the spectrum rather than the slope. This suggests that color temperature calibrated RGB channels can be corrected for relative differences in atmospheric scattering and absorption to allow for quantitative intercomparison. A mosaic of 18 intercalibrated RGB photos renders a spectral feature space with four clearly defined spectral endmembers corresponding to white, yellow and red light sources, with brightness modulated by a dark background endmember. These four spectral endmembers form the basis of a linear spectral mixture model which can be inverted to provide estimates of the areal fraction of each endmember present within every pixel instantaneous field of view. The resulting spectral feature space shows two distinct mixing trends extending from the dark endmember to near flat spectrum (white-yellow) and warm spectrum (orange) sources. The distribution of illuminated pixels is strongly skewed toward a lower luminance background of warm spectrum street lighting with brighter lights generally corresponding to point sources and major thoroughfares. Intercomparison of 18 individual urban lightscape spectral feature spaces show consistent topology, despite variations in exposure and interior mixing trends.

## Introduction

The widespread and increasing use of satellite observations of anthropogenic night light for a variety of applications, ranging from socioeconomic to ecologic is a testament to the unique information content of these data. Whereas the DMSP-OLS series of meteorological satellites was the only source of satellite night light data for many years, there have been more than 10 satellites with nocturnal imaging capability launched since 2000 *(Levin et al. 2020)*. However, none of the non-commercial night light sensors provide both sub-hectometer spatial resolution and multispectral visible bands. Sub-hectometer spatial resolution is necessary to resolve the urban light field at the decameter characteristic scale of the urban mosaic while multispectral (or preferably hyperspectral) imaging is necessary to distinguish the diversity of light sources used to illuminate the built environment. The importance of the color spectrum of lighting sources has been recognized for its potential impact on ecosystems *(Rich and Longcore 2013; Gaston et al. 2012; Gaston et al. 2015) (Davies et al. 2013; Hölker et al. 2010)*, astronomical light pollution *(Cinzano, Falchi, and Elvidge 2001) (Cinzano and Falchi 2012)*, melatonin suppression *(Brainard et al. 2001)* and as an epidemiological correlate to multiple types of cancer *(Garcia-Saenz et al. 2018; Garcia-Saenz et al. 2019)*.

While future night light sensors will likely incorporate multispectral low light imaging capability, there remains only one option for sub-hectometer multispectral night light imaging with extensive geographic coverage over the past two decades: Astronaut photos taken from the International Space Station (ISS). At the time of writing, NASA's Gateway to Astronaut Photography of Earth (*https://eol.jsc.nasa.gov*) contains over 30000 photos of ~2500 cataloged urban lightscapes taken from the ISS. There is considerable redundancy, as many of these photos are multi-shot sequences from slightly differing view geometries as the camera's field of view changes slightly due to spacecraft motion. In addition, many are panoramic photos of large regions containing multiple cities with insufficient resolution for many scientific applications.

Despite these limitations, a subset of the collection of astronaut photos of urban lightscapes are of sufficient spatial resolution, sharpness and exposure to be potentially useful for spectral characterization of urban lightscapes. Particularly as no other geographically extensive source of comparable imagery is available for the past two decades. Variations in location, date, view geometry, resolution and exposure impose limitations to comparative analyses, but with cross calibration it is possible to obtain a spectral characterization of the diversity of light sources contributing to urban lightscapes over the past decade. Spectral characterization of multiple urban lightscapes can provide a basis for quantifying intra and interurban variability in night light brightness and extent. However, the disparity between the centimeter scale of most light sources and the decameter spatial resolution of even the most detailed of astronaut photos implies the potential for subpixel spectral mixing to occur in the sensor's Instantaneous Field of View (IFOV) of individual pixels.

The objective of this study is to characterize the spectral properties of a diverse collection of urban lightscapes. Specifically, to quantify the luminance and color distribution of the spectral feature space of multispectral night light. In order to accommodate the spectral mixing of multiple light sources within individual image pixels, a spectral mixture analysis is conducted on a mosaic of intercalibrated color images to identify spectral endmembers bounding the spectral feature space. The spectral endmembers form the basis of a linear spectral mixture model that can be inverted to yield endmember fraction estimates for each RGB image pixel. The trivariate distributions of endmember fractions quantify the spectral diversity of individual urban lightscapes as well as the aggregate of the full mosaic.

## Data

All color images were obtained as Nikon raw files (.nef) from NASA's Gateway to Astronaut Photography of Earth (*https://eol.jsc.nasa.gov*). The .nef raw files were converted to digital negative (.dng) format using Online Converter (*https://www.onlineconverter.com/nef*), imported to Adobe Camera Raw v4.0, standardized to common brightness temperatures (2200 K & 5500 K; 0 tint) and saved as 32 bit RGB uncompressed TIFF (.tif) images. Linear tone curves were retained with default settings for brightness (+50) and contrast (+25). Default settings were also retained for sharpening (+25) and noise reduction (luminance: 0; color: +25). All subsequent analysis was performed using ENVI and IDL software.

A total of 122 high quality photographs were considered for analysis. All were shot with Nikon D3S, D4 or D5 cameras. The majority were shot with either a Nikkor 400 mm f/2.8D IF-ED lens (67 photos) or a Nikkor 180 mm f/2.8 AF-D lens (26 photos). A comparison of the effects of view geometry and lens focal length (hence image dimensions and ground sample distance) is shown in Figure 1. The primary photo selection criteria were high spatial resolution and minimal blur. Blur can result from either camera movement during exposure or atmospheric water vapor scattering as illustrated in Figure 2. The tradeoff between image sharpness and background noise level of the photos is determined by both exposure time and ISO setting, as well as noise reduction and sharpening settings used in raw conversion. Almost all photos were acquired at maximum aperture (f/2.8 for both lenses). As shown in Figure 3, most photos were acquired with exposures in the 1/4 to 1/60 second range with ISO settings near 10000. From this set of 122 candidate photos, 18 were chosen on the basis of image quality, spatial resolution and geographic diversity. City names (as provided by NASA), dates and image IDs are given in Table 1.

Effects of atmospheric absorption and scattering were compensated using the MODTRAN radiative transfer model *(Berk et al. 2014)* to simulate atmospheric transmittance for different model atmospheres. Modeled transmittance profiles were convolved with spectral response functions for the Nikon D3S *(Metcalf 2012)* to yield weighted transmittance estimates for the R, G and B image channels. Spectral responses are very similar for the D3S, D4 and D5 sensors *(Sánchez de Miguel et al. 2019)*, with the exception of increased Near Infrared sensitivity for the D5. Only London and Las Vegas were acquired with the D5. As shown in Figure 4, molecular absorptions are more pronounced for the Tropical than for the Mid-Latitude Winter model atmosphere, but the overall amplitude and curvature of the transmittance profiles are nearly identical. Amplitude and curvature are far more sensitive to the visibility (Aerosol Optical Depth; AOD) than model atmosphere, as indicated in Figure 4. In principle, it should be possible to obtain location-specific visibility estimates from the Aerosol Robotic Network (AERONET) database (https://aeronet.gsfc.nasa.gov). Unfortunately, time and location-coincident estimates were not available for any of the 18 sites chosen for this analysis, so a common visibility of 23 km (MODTRAN default) was used for all corrections. Channel-specific correction terms are obtained from the complement of the response-integrated transmittance and added to the unit-normalized DN value of each image pixel.

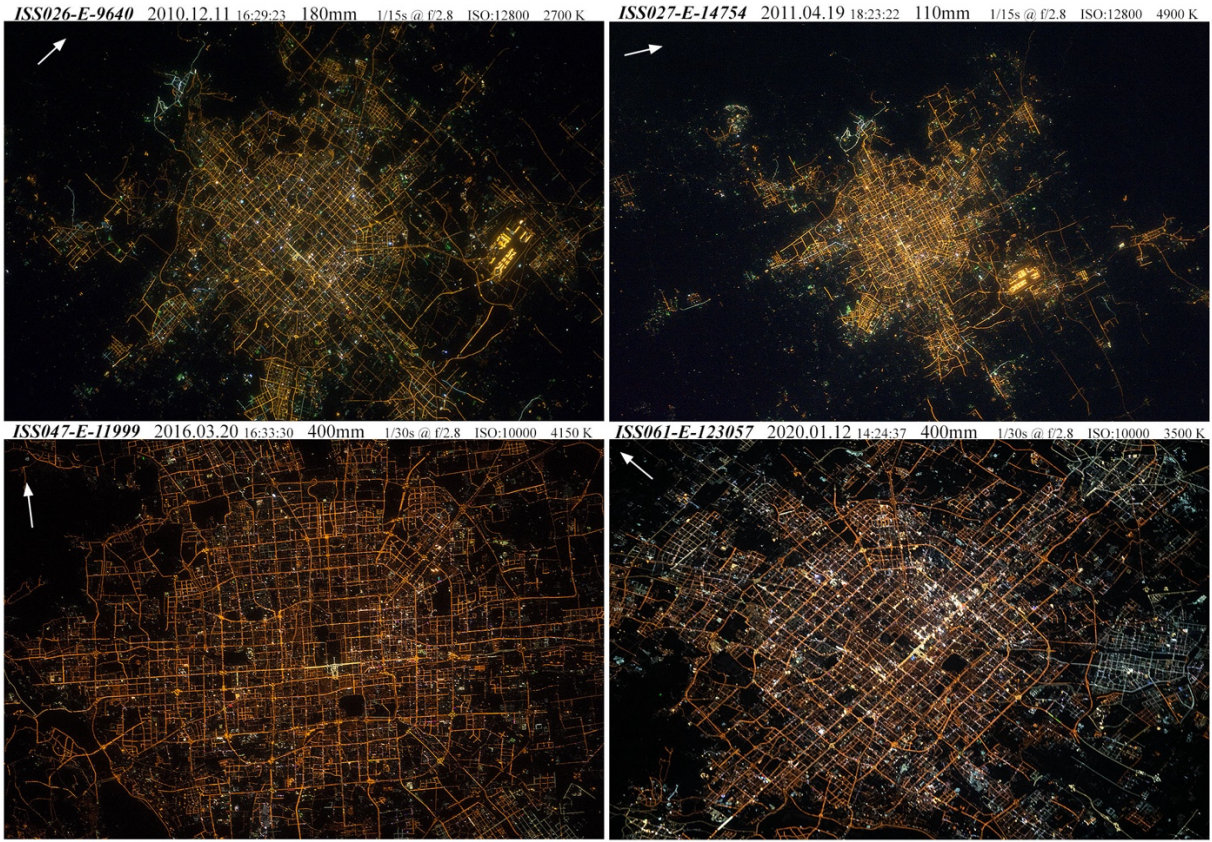

*Figure 1  Urban lightscape comparison of Beijing photographed at different dates, local times, view geometries, focal lengths, exposures and white balance settings.  The 2010 and 2011 shots (top) were taken with the same camera (Nikon D3S) less than 4 months apart and differ primarily in lens focal length (hence spatial resolution) and white balance temperature.  The 2016 and 2020 shots (bottom) were taken with different cameras (Nikon D4 & D5) using the same lens, exposure and ISO and similar white balance temperature, but different local times and view geometries. The greater number and brightness of white lights within the four inner ring roads in the 2020 shot may be a combined result of the earlier local time and more oblique view geometry imaging more illuminated facades and commercial lighting not seen in the near-nadir view shot taken after midnight local time.  Arrows in UL corners show ~North.*

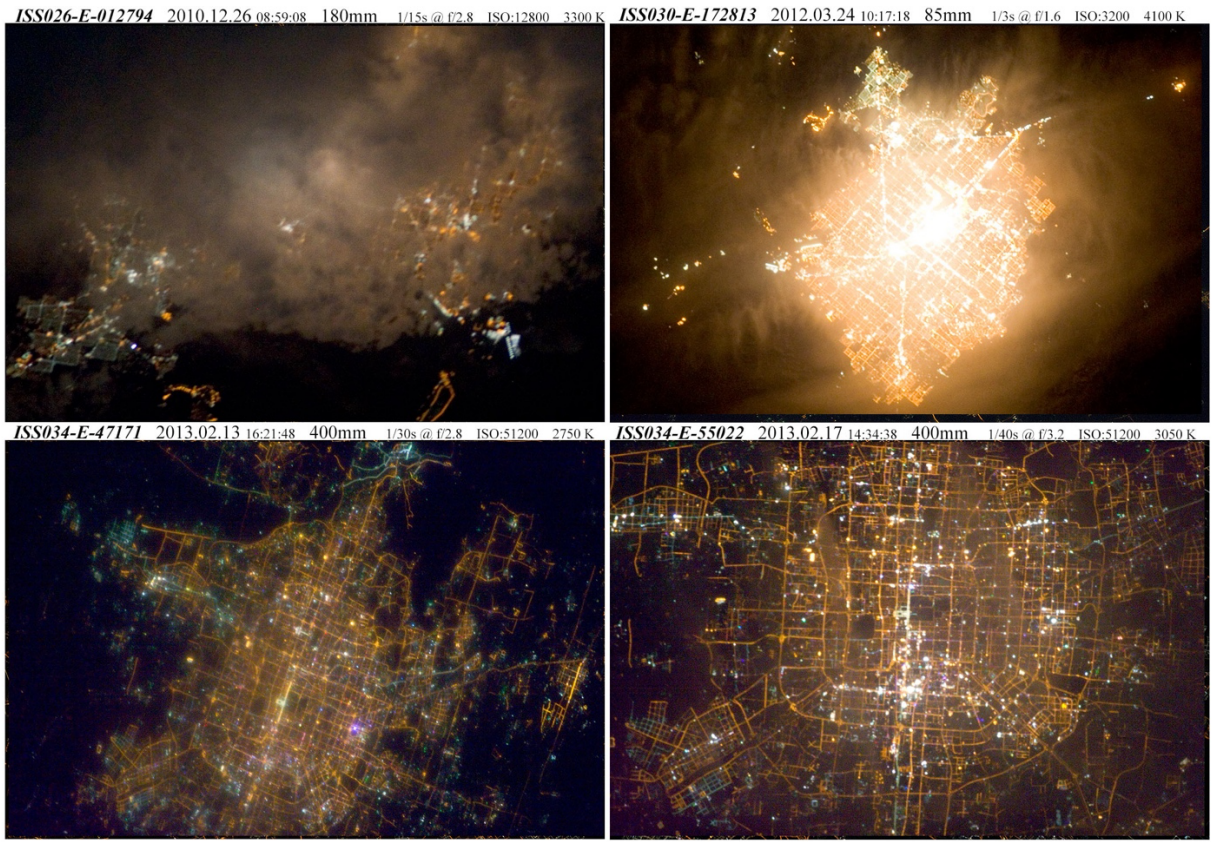

*Figure 2   Spatially variable atmospheric scattering effects for Las Vegas (top) and Beijing (bottom).  Translucent clouds over Las Vegas are more conspicuous because of overglow effects extending beyond the periphery of the light sources.  Spatially varying sharpness within the lighted area of Beijing is more subtle but nonetheless distorts both brightness and spatial extent of individual light sources.  Compare the sharpness of these Beijing images with the 2016 and 2020 shots shown in Figure 1.   All 4 Beijing images were taken with the same 400mm lens.*

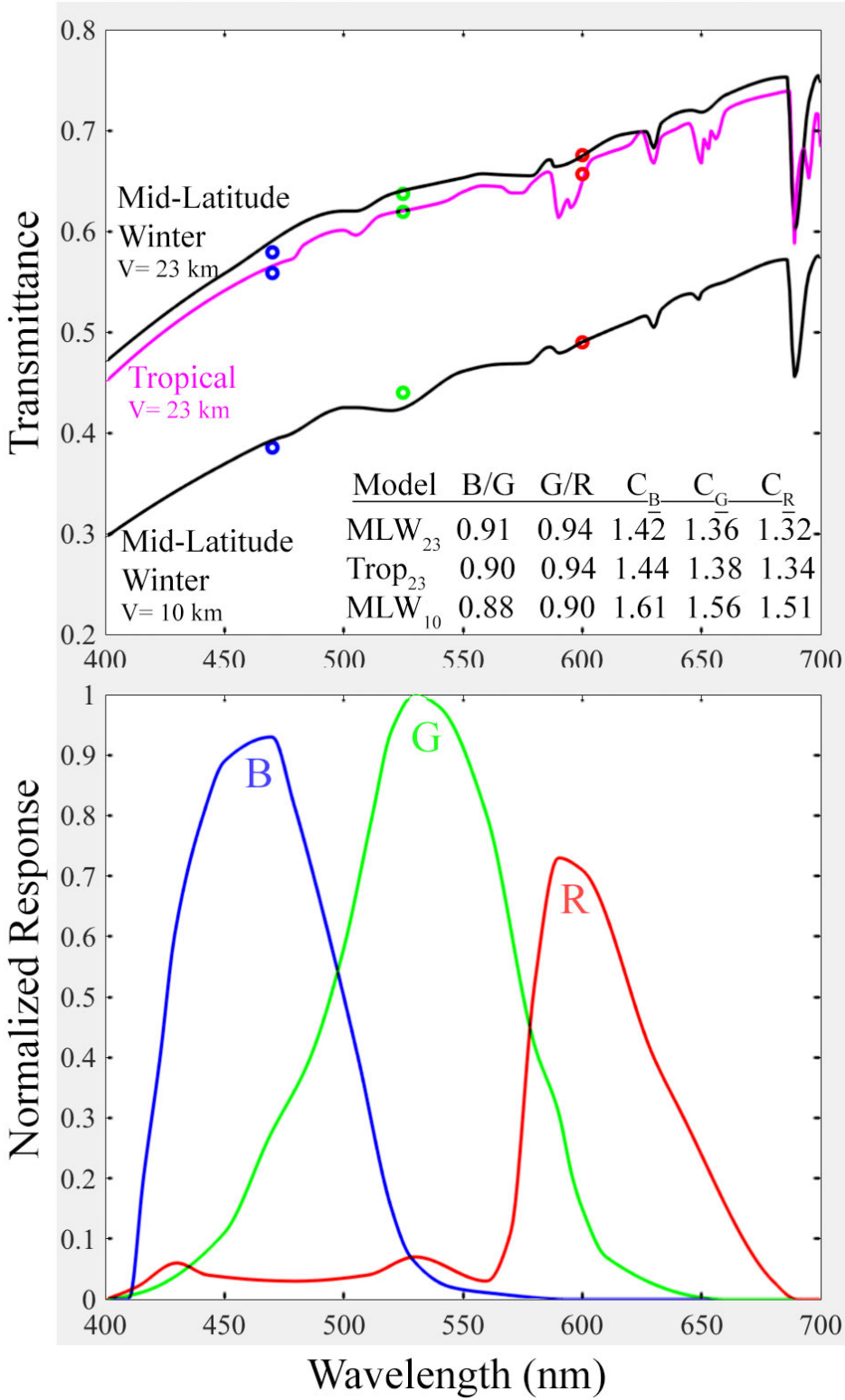

*Figure 3  Atmospheric transmittance correction estimation.  Nikon D3S spectral responses (bottom) are convolved with MODTRAN-derived atmospheric transmittance (top) for different visibilities and atmosphere models to produce response-weighted channel-specific estimates (circles) for transmittance loss corrections ($C_{BGR}$).*

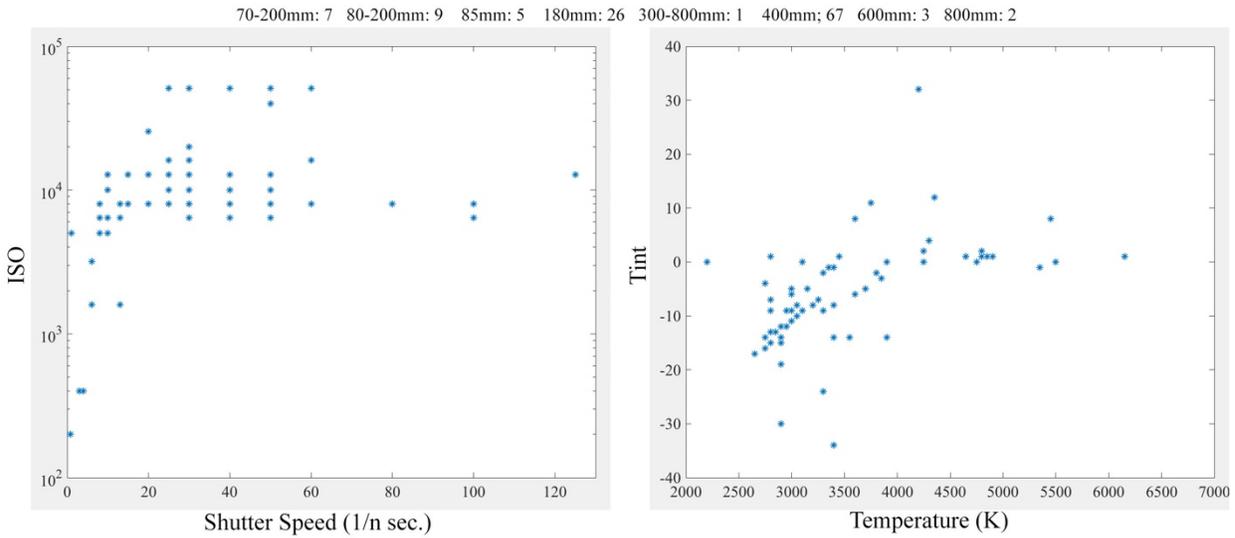

*Figure 4 Exposure, ISO and white balance settings for 122 high quality urban lightscape photos taken between 2003 and 2020. Most photos were taken with high ISO and relatively low shutter speed for the telephoto focal lengths of the lenses used (distribution at top). The distribution of white balance settings is skewed toward low temperatures consistent with the widespread use of high and low pressure Sodium light sources.*

Table 1

| Name | ISS ID | Date:Time | Tilt | Alt (km) | F (mm) |
|---|---|---|---|---|---|
| **Antwerp** | iss035e017345 | 2013.04.07:00:12:44 | 07°E | 396 | 400 |
| **Bangkok** | iss046e000169 | 2015.12.12:17:05:11 | 45°W | 404 | 400 |
| **Beijing** | iss026e009640 | 2010.12.11:16:29:23 | 99°SW | 343 | 180 |
| **Berlin** | iss035e017202 | 2013.04.06:22:37:22 | 28°NE | 398 | 400 |
| **Calgary** | iss045e155033 | 2015.11.28:07:07:47 | --/-- | 394 | 400 |
| **Chicago** | iss047e043884 | 2016.04.05:06:17:11 | --/-- | 398 | 400 |
| **HoChiMinh** | iss046e000196 | 2015.12.12:17:06:12 | --/-- | 404 | 400 |
| **Istanbul** | iss032e017547 | 2012.08.09:23:38:20 | 45°SW | 396 | 400 |
| **Kuwait** | iss032e017635 | 2012.08.09:23:43:31 | 41°SW | 398 | 400 |
| **LasVegas** | iss062e061134 | 2020.02.27:11:06:14 | 26°NW | 415 | 400 |
| **London** | iss061e052957 | 2019.11.20:20:41:38 | 53°NE | 413 | 400 |
| **LosAngeles** | iss026e006228 | 2010.11.30:12:04:22 | 45°SW | 350 | 180 |
| **Naples** | iss032e014256 | 2012.08.05:20:40:36 | 42°SW | 396 | 400 |
| **NewYork** | iss026e008537 | 2010.12.08:06:08:04 | 99°W | 346 | 400 |
| **Mecca** | iss034e51161 | 2013.02.17:20:52:14 | 32°W | 413 | 400 |
| **Paris** | iss043e093480 | 2015.04.08:23:18:37 | 17°NE | 394 | 400 |
| **Phoenix** | iss035e005438 | 2013.03.16:11:56:50 | 14°SE | 396 | 400 |
| **Singapore** | iss041e004915 | 2014.09.13:18:48:49 | 36°W | 417 | 800 |

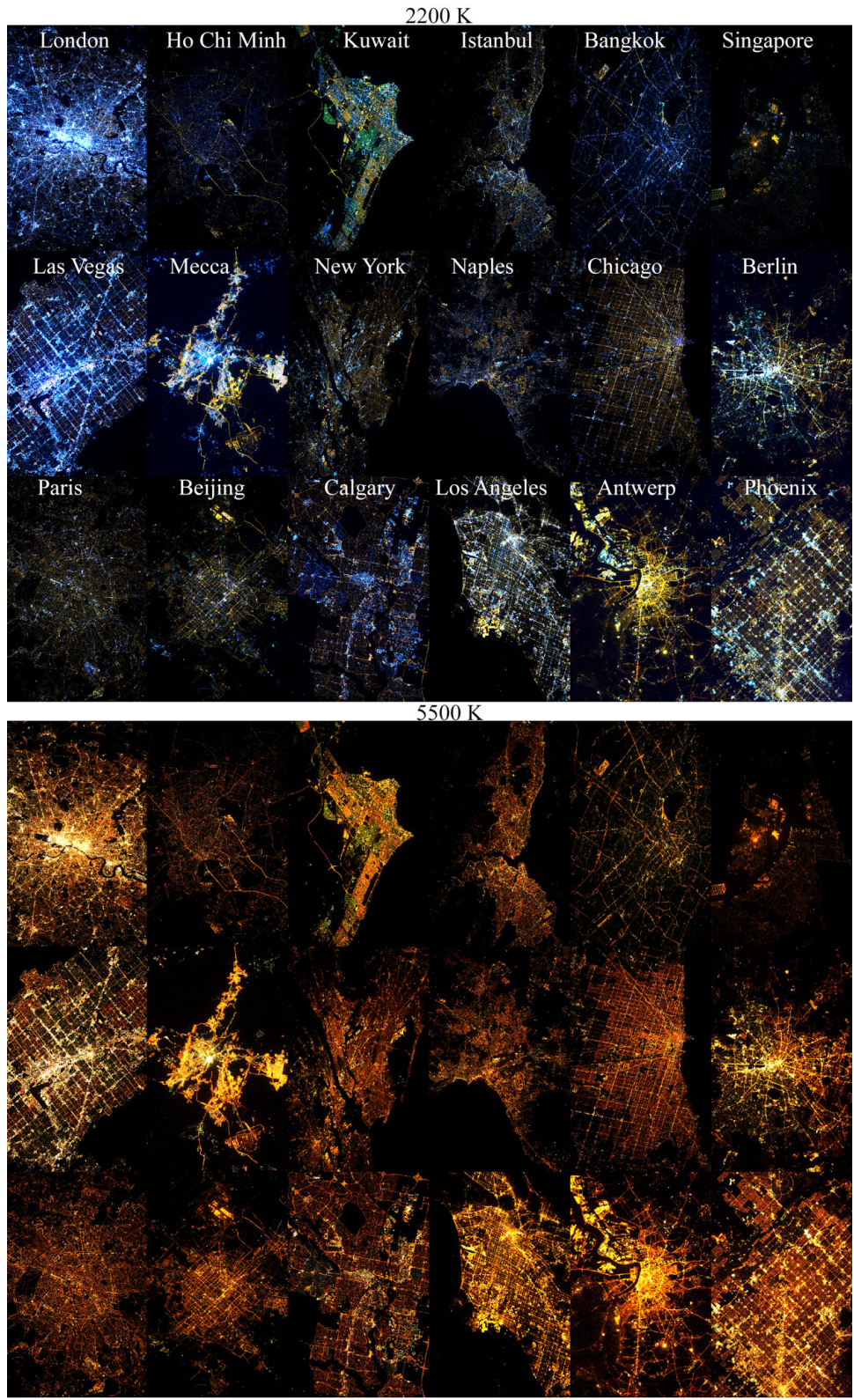

*Figure 5 Color temperature-calibrated mosaics of 18 cities. The 2200 K mosaic corresponds to the color temperature of high pressure Sodium lamp while the 5500 K mosaic corresponds to natural daylight - showing the preponderance of warm spectrum light sources.*

## Methods

Originally developed for spectroscopic analysis of lunar and martian substrates, the linear spectral mixture model assumes that spectrally distinct materials within the sensor Instantaneous Field of View (IFOV) contribute to an aggregate mixed reflectance (or radiance) spectrum in proportion to their relative areal fraction of the IFOV *(Singer and McCord 1979; Adams, Smith, and Johnson 1986; Johnson, Smith, and Adams 1985; Johnson et al. 1983)*. Given N spectrally distinct endmember spectra, and a D (> N) channel multispectral image, a linear spectral mixture model can be given as a system of D channel-specific linear mixing equations, each containing N terms to represent the areal fraction of each spectral endmember in the observed spectrum. In matrix notation: $CF = O$, where $C$ is the NxD matrix of endmember reflectances, $F$ is the vector of endmember fractions to be estimated and $O$ is the observed spectrally mixed pixel vector. If D > N the system is overdetermined, allowing for the possibility of a unique, or at least optimal, solution. Inversion of the linear model yields an estimate of the fraction of each endmember possibly present in the observed mixed spectrum.

The spectral endmembers on which the linear mixture model is based may be chosen *a priori* from field or laboratory spectra, or from a spectral mixture analysis of all observed spectra to be modeled *(Adams and Gillespie 2006; Smith et al. 1990)*. While there exist a multitude of ways to select spectral endmembers, all require either assumptions about how many and which endmembers may be present, or a characterization of the observed spectra to identify endmembers relative to all spectra available. In the latter case, some form of dimensionality reduction is generally applied to the D dimensional feature space of observed spectra. If the topology of the lower dimensional projection of the spectral feature space suggests linear mixing, a convex hull can be circumscribed bounding all or most of the observed spectra in the projection. In the projection, the endmembers reside at the apexes of the space, with binary mixtures occurring along the line segments between each adjacent pair of endmembers and N endmember mixtures occurring within the bounding hull. *(Boardman 1990; Boardman 1989; Boardman 1993)*.

With multispectral land surface reflectance, large spectrally diverse collections of spectra can provide an approximation of the full global feature space *(Small 2004)*. Identification of spectral endmembers bounding the composite feature space provides a basis for a general mixture model for land cover reflectance. Sensor-specific standardized endmembers identified from such compilations indicate that feature spaces for Landsat *(Small and Milesi 2013; Sousa and Small 2017)*, MODIS *(Sousa and Small 2019)* and Sentinel 2 *(Small 2018)* all have similar mixing space topology bounded by common spectral endmembers representing rock and soil substrates, green vegetation and water (for ice free landscapes). This allows for identification of sensor-specific standardized spectral endmembers upon which standardized spectral mixture models may be based. Inversion of the linear spectral mixture model using standardized spectral endmembers is effectively a change of basis from reflectance to land cover fraction. This change of basis from higher dimensional reflectance to lower dimensional land cover fraction renders continuous variations in the landscape interpretable in the context of the most distinct physical properties impacting its form and function.

Because anthropogenic light sources (e.g. bulbs, tubes and LEDs) have scales on the order of centimeters, and the IFOV of the camera pixels have scales on the order of decameters (at ISS altitudes), we can treat the radiance incident on the camera sensor as a spectral mixture of multiple light sources coming from a combination of direct and reflected light within the pixel IFOV. While some pixels may be dominated by a single light source, the most general case can include multiple light sources with both direct illumination and upward reflected light. Thus, the strategy is to characterize the spectral distribution of night light color imaged from a variety of urban lightscapes to derive standardized spectral endmembers for visible night light. Following the approach described by *(Small 2004)*, a spectrally diverse mosaic of 18 urban lightscapes is constructed from MODTRAN-calibrated photos. For comparison, mosaics are constructed for two different white balance calibrations. A 2200 K calibration corresponds to the color temperature of high pressure Sodium lamps widely used for urban street lighting, while a 5500 K calibration corresponds to the color temperature of sunlight. Both mosaics are shown for comparison in Figure 5. The daylight-relative 5500 K calibration illustrates the preponderance of warm spectrum lighting used in large areas of all 18 lightscapes, while the warm spectrum calibration of 2200 K shows greater contrast within and among lightscapes. All subsequent analyses are performed on the 2200 K calibrated mosaic.

The spectral feature space of the lightscape mosaic allows for identification of the number and identity of spectral endmembers, the linearity of spectral mixing, and the trivariate distribution of the entire set of illuminated pixels. A principal component transform is applied to identify the apexes corresponding to spectral endmembers and to characterize linearity of spectral mixing. Figure 6 clearly shows four apexes bounded by binary linear mixing lines. Because the principal component transform maximizes variance, the first PC is analogous to overall luminance while the second and third PCs reveal a planar triangular mixing space bounded by the three brightest spectral endmembers representing white, yellow and red sources (lower right). Note that very few pixels are fully red with the majority clustering interior of the red apex and having a spectrum closer to orange than red. The 3D spectral feature space can be considered a tetrahedral pyramid with the brightest white, yellow and red sources forming a triangular base with a gray axis extending to the dark apex.

The White, Yellow, Red and darK endmembers span the feature space and therefore form the basis of the WYRK linear mixture model. However, the number of RGB color channels (3) is less than the number of endmembers (4), rendering the problem underdetermined. This can be resolved by adding a unit sum constraint equation ($F_1+F_2+F_3+F_4 = 1$) to the system. In this analysis the linear mixture model is inverted using the familiar least squares solution: $(C^TC)^{-1}C^T$ *(Settle and Drake 1993)* to invert the 4 endmember linear mixture model for each RGB vector in the lightscape mosaic. The resulting distribution of fraction estimates is generally well-bounded [0,1] for each endmember, with a small number (912/5000000 ~ 0.018%) of blue light sources having slightly negative values for the Yellow endmember fraction. The distribution of dark fraction estimates has 95% of pixel spectra with dark fractions $> 0.78$. As expected for such a strongly linear mixing space, the RMS misfit to the 4 endmember linear model is less than $10^{-7}$ DN for all pixel spectra.

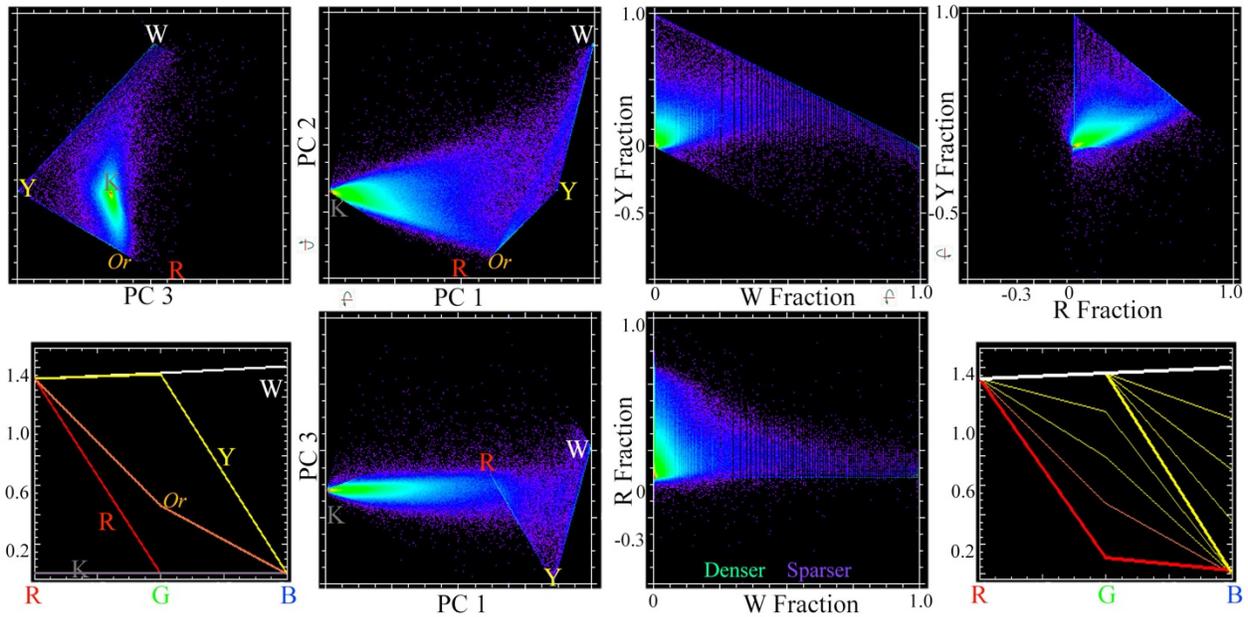

*Figure 6  3D spectral feature space and spectral fraction space for the 18 city composite (2200 K).  Density-shaded orthogonal projections of the principal component distributions (left) show luminance corresponding to PC 1 with White, Yellow and Red endmembers bounding a triangular plane of maximum luminance perpendicular to the gray axis extending to the dark (K) endmember.  Since very few pixels are pure red, the distribution trends toward orange mixtures near the red endmember.   Projections of the WYRK fraction space (right) are largely bounded by linear mixtures with a small number of blue pixels having slightly negative yellow or red fractions.*

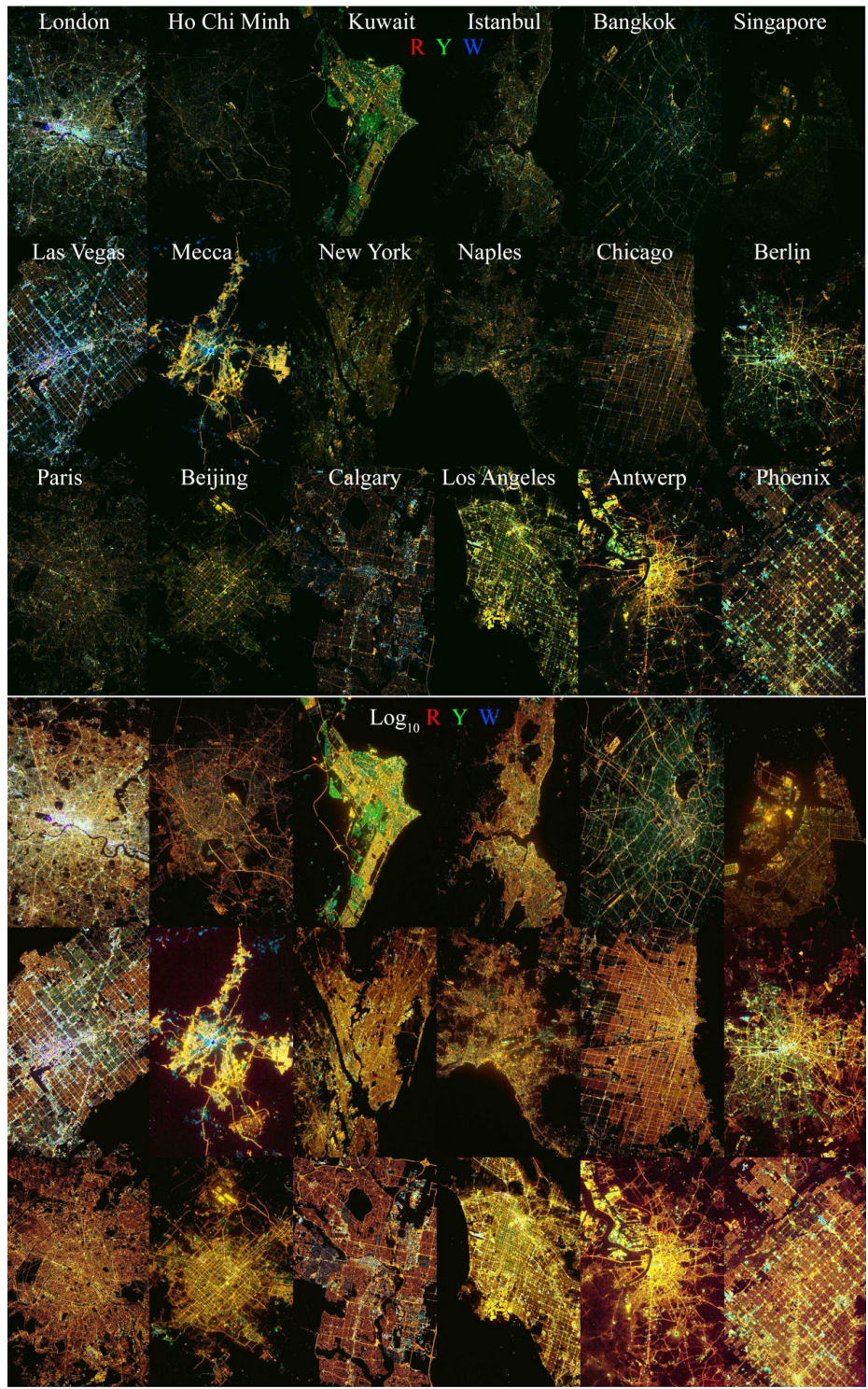

*Figure 7 Spectral endmember fraction mosaics of 18 cities. The 2200 K mosaic is unmixed with a 4 endmember linear mixture model with White, Yellow, Red and Dark endmembers. Because the Dark endmember fraction modulates luminance, a RGB composite of the R, Y and W fractions (top) resembles the 2200 K RGB composite in Fig. 5. The RGB composite of Log$_{10}$ (R, Y, W) (bottom) partially offsets this Dark fraction modulation to enhance the lower luminance mixed spectra. Both mosaics displayed with a 1% linear stretch applied.*

# Results

Because the dark endmember fraction simply modulates brightness as the complement to the White, Yellow and Red endmember fractions, the essential information content of the WYRK model can be conveyed with an RGB composite of the WYR fractions. When [R, G, B] = [R, Y, W], the resulting composite maintains a comparable color composition to the 2200 K calibrated RGB mosaic. Compare Figure 7 (top) to Figure 5 (top). The WYR composite shows a somewhat more uniform brightness distribution as the transformation to endmember fraction offsets the greater brightness of whiter light sources in most of the lightscapes. This is most apparent in the greater spectral diversity of the London and Las Vegas lightscapes in the WYR composite compared to the 2200 K RGB composite. Another benefit of the WYR model is that pixels saturated in a specific endmember appear either red, green, blue, cyan magenta or yellow to indicate which endmembers are saturated.

One consequence of the equalizing effect of the endmember fraction basis is to skew the distribution of fractions heavily on the white fraction axis corresponding to overall brightness. This is apparent in the slightly darker cast of the WYR composite in Figure 7 compared to the 2200 K RGB composite in Figure 5. The extent of the dimmer more diffuse light sources can be emphasized by displaying $Log_{10}$ of WYR fractions, as shown in Figure 7. The primary spectral effect of this brightening is to emphasize the large areas of diffuse warm spectrum light.

Bivariate fraction distributions from orthogonal projections of the WYRK feature space are shown for 8 contrasting urban lightscapes in Figure 8. While each individual lightscape feature space has topology consistent with the composite space, some differences are apparent. Primarily in the relative abundance of saturated pixels on the binary mixing line spanning the white and yellow endmembers. This is most apparent for London, Berlin and Las Vegas. There is also some orange saturation near the red endmember on the binary mixing line spanning red and yellow. This is most apparent for Mecca and Los Angeles. In addition, there are a few examples of cool spectrum light sources with negative fractions of the yellow endmember. This is most apparent in London, Berlin and Las Vegas. The considerable variation in the skewness of the fraction spaces, most apparent in the white-yellow projection, is a result of differences in actual brightness distributions, atmospheric opacity and image exposure.

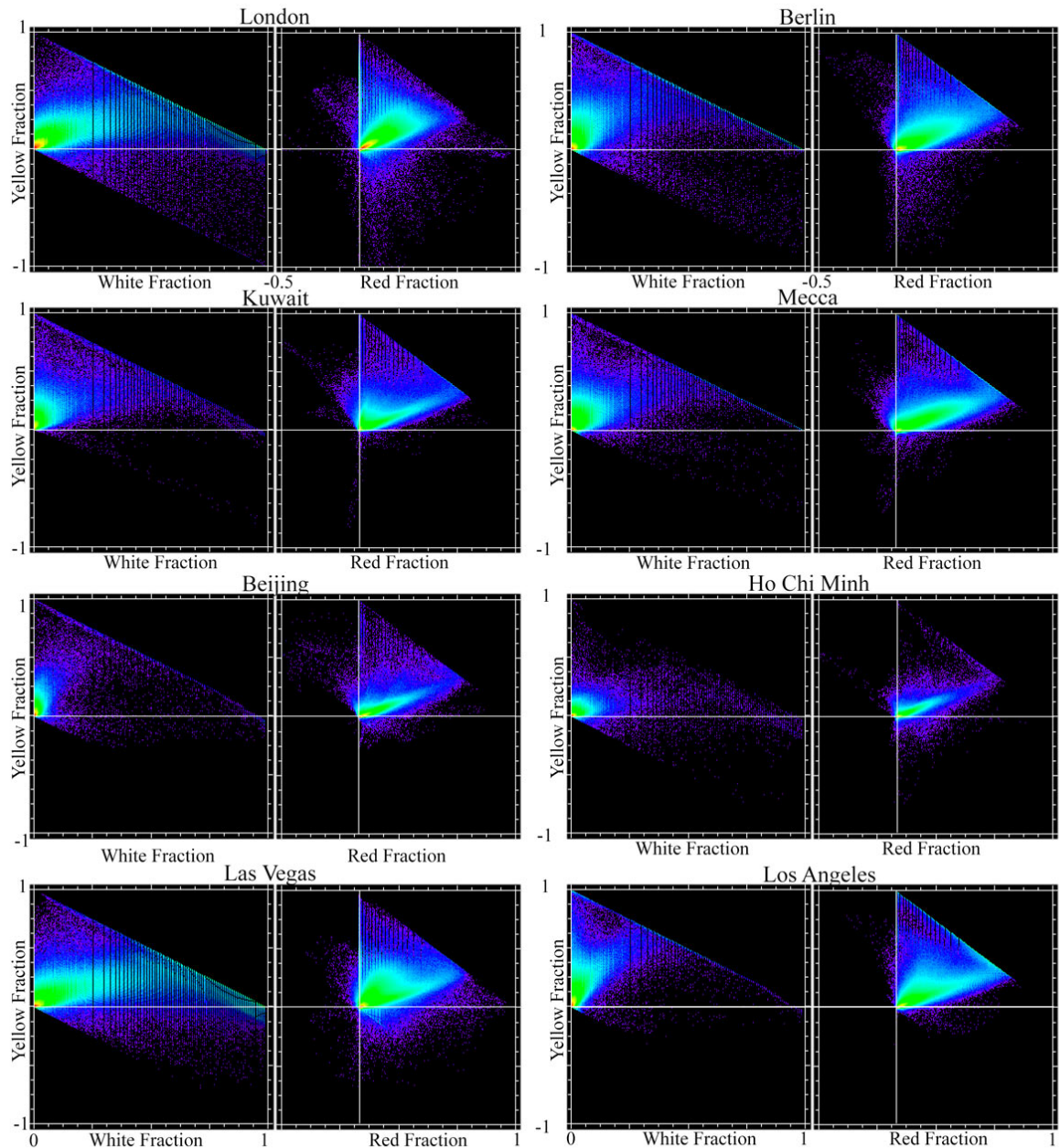

*Figure 8 WYR fraction spaces for 8 contrasting lightscapes. The white-yellow projections show varying degrees of saturation on the binary mixing line related to varying exposures of different photographs. Regardless of exposure, all distributions are strongly skewed toward dimmer lights (near the origin) with distinct continua extending to a warm spectrum orange near the red endmember on the yellow-red projection.*

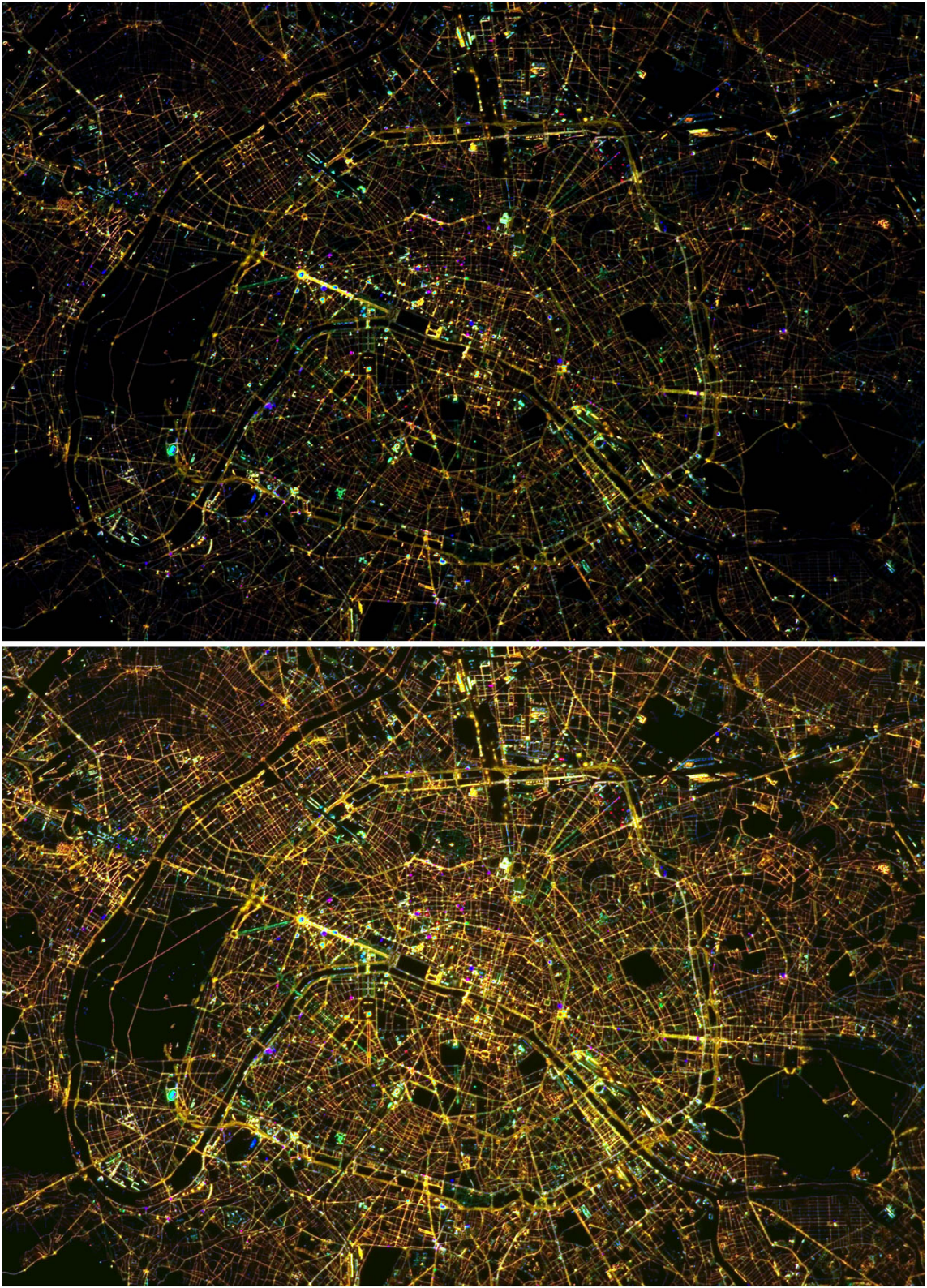

*Figure 9 WYR and Log$_{10}$WYR fraction composites for Paris illustrate the diversity of high luminance colored point sources superimposed on the pervasive low luminance warm spectrum street lighting.*

**Discussion**

Implicit in this analysis is the assumption that the varying exposures of the individual photos will partially compensate for differences in atmospheric path length and aerosol optical depth among photos, making a comparison of the relative RGB differences meaningful. The intercalibration and transmittance correction to relative RGB values used here is far simpler than the absolute calibration procedure proposed *by (Sánchez de Miguel et al. 2021),* however absolute radiances are not required for the spectral mixture analysis described here. The comparison of modeled MODTRAN transmittances clearly indicates that the primary effect of both path length and AOD is to shift the bias of the transmittance without significantly affecting the slope – suggesting that the wavelength dependent Rayleigh scattering is the primary influence on relative RGB values. If this is indeed the case, then intercalibration to a common color temperature and compensation for relative transmission losses among R, G and B channels should allow for intercomparison of the RGB spectral feature spaces. The consistency of the overall topology of the most spectrally distinct feature spaces shown in Figure 8 suggests that this is the case.

The unavoidable aliasing of narrowband lamp spectra by broadband RGB sensors will render many different light sources indistinguishable. The comparison of G/R and B/G ratios given by *(Sánchez de Miguel et al. 2019)* for 50 common street lamp spectra forms a continuum ranging from warm spectrum low pressure Sodium and LED to nearly flat spectrum ceramic metal halide, fluorescent and LED. Only green fluorescent and LED deviate from this continuum. Even in a luminously diverse environment like the Las Vegas Strip, a spectral mixture analysis of 360 channel SpecTIR hyperspectral night light imagery by *(Small et al. 2018)* found the spectral feature space of the brightest lights to be effectively 4 dimensional spanned by high pressure Sodium, incandescent, and two distinct metal halide spectra. In addition, *(Kruse and Elvidge 2011)* and *(Metcalf 2012)* also identified red and blue neon spectra from the same hyperspectral data. In contrast, the photograph of Las Vegas used in this analysis shows two distinct continua extending from the dark endmember to a warm spectrum orange source (presumably low pressure Sodium) and a saturated nearly white source that may conflate a variety of bright sources that were overexposed in the ISS photograph.

In spite of the saturation resulting from varying exposures of different photos, the feature spaces of the 8 individual lightscape examples in Figure 8 are self-consistent relative to the feature space of the full mosaic of 18 lightscapes. Despite varying degrees of exposure evident in the white-yellow projection of the space, all have similar continua extending to an orange source near the red endmember in the yellow-red projection. This very likely reflects the widespread use of high and low pressure sodium lamps for street lighting. This is apparent in the $Log_{10}$ transformed mosaic where these warm spectrum sources are seen to be widespread throughout all 18 cities – including those with significant areas of cooler spectrum lighting (e.g . Calgary, Kuwait, Las Vegas and London). While cooler spectrum LED streetlights are replacing warm spectrum Sodium lamps in many urban areas, there is apparently strong controversy related to both the color cast of warm spectrum Sodium lamps (e.g. *(Moser 2017)*) as well as the brightness of LED lamps *(Stark 2018)* in some cities. Despite the aforementioned aliasing of lamp spectra by broadband RGB sensors, the photos used in this analysis clearly distinguish warm and cool spectrum mixing trends within the individual lightscape feature spaces.

A full resolution comparison of the WYR and Log$_{10}$WYR composites for Paris illustrates several general features apparent in the 18 lightscapes analyzed here, and many of the other 122 with sufficient resolution and sharpness to resolve decameter-scale features.  A can be seen on Figure 9, the brightest lights tend to be point sources, and sometimes large thoroughfares like the Avenue de Champs-Élysées and the Boulevard Périphérique.  These point sources are often high luminance white, or saturated on the yellow-white mixing line, however, the Paris image shows a variety of different colors for isolated point sources.  In contrast, the lower luminance pixels that form the bulk of the distribution are warm spectrum street lighting on the dark-orange continuum.  As most street lights are designed to project light downward, the lower luminance warm spectrum distribution is expected to be primarily upward scattering from street light reflected from ground surfaces and building sides, while the brighter point sources would include upward directed lighting on building facades that may be more apparent in oblique orientation photos.  Comparisons with calibrated VIIRS Day Night Band imagery indicates that this reflected background luminance is 1 to 2 orders of magnitude dimmer than the brightest point sources that are able to dominate even a hectometer-scale IFOV like VIIRS DNB (~700 m) *(Small 2019)*.

In summary, the NASA archive of astronaut photos of urban lightscapes currently contains more than 100 decameter resolution RGB images of sufficient quality to allow for spectral mixture analysis of the areal distribution of different color light sources from almost 100 cities around the world.  While intercomparisons of absolute radiance are complicated by variations in view geometry, atmospheric path length, exposure and aerosol optical depth, a relatively simple standardization of color temperature and correction for relative transmissivity of red, green and blue image channels can allow for comparative spectral mixture analysis of the relative luminance distributions and spectral feature spaces of these urban lightscapes, and possibly for multitemporal change analyses on longer time scales.

## References

Adams, J.B., and A.R. Gillespie. 2006. *Remote Sensing of Landscapes with Spectral Images* (Cambridge University Press: Cambridge, UK).

Adams, J.B., M.O. Smith, and P.E. Johnson. 1986. 'Spectral mixture modeling; A new analysis of rock and soil types at the Viking Lander 1 site', *Journal of Geophysical Research*, 91: 8098-122.

Berk, A., P. Conforti, R. Kennett, T. Perkins, F. Hawes, and J. van den Bosch. 2014. "MODTRAN6: a major upgrade of the MODTRAN radiative transfer code." In *Proc. SPIE 9088, Algorithms and Technologies for Multispectral, Hyperspectral, and Ultraspectral Imagery XX*. SPIE.

Boardman, J. 1990. 'Inversion of high spectral resolution data', *SPIE - Imaging Spectroscopy of the Terrestrial Environment*, 1298: 222-33.

Boardman, J. W. 1989. "Inversion of imaging spectrometry data using singular value decomposition." In *IGARSS'89 12th Canadian Symposium on Remote Sensing*, 2069-72. Vancouver, B.C.

Boardman, J.W. 1993. "Automating spectral unmixing of AVIRIS data using convex geometry concepts." In *Fourth Airborne Visible/Infrared Imaging Spectrometer (AVIRIS) Airborne*

*Geoscience Workshop*, edited by R.O. Green, 11-14. Jet Propulsion Laboratory, Pasadena, CA.

Brainard, G.C., J. Hanifin, J. Greeson, B. Byrne, G. Glickman, E. Gerner, and M.D. Rollag. 2001. 'Action spectrum for melatonin regulation in humans: evidence for a novel circadian photoreceptor. ', *Journal of Neuroscience*, 21: 6405–12.

Cinzano, P., and F. Falchi. 2012. 'The propagation of light pollution in the atmosphere', *Monthly Notices of the Royal Astronomical Society*, 427: 3337–57.

Cinzano, P., F. Falchi, and C.D. Elvidge. 2001. 'The first world atlas of the artificial night sky brightness', *Monthly Notices of the Royal Astronomical Society*, 328: 689-707.

Davies, T. W., J. Bennie, R. Inger, N. Ibarra, and K. Gaston. 2013. 'Artificial light pollution: are shifting spectral signatures changing the balance of species interactions?', *Global Change Biology*, 19: 1417–23.

Garcia-Saenz, A., A. Sánchez de Miguel, A. Espinosa, A. Valentin, N. Aragonés, J. Llorca, P. Amiano, V. Martín Sánchez, M. Guevara, R. Capelo, A. Tardón, A. Peiró-Perez, J. Jiménez-Moleón, A. Roca-Barceló, B. Pérez-Gómez, T. Dierssen- Sotos, T. Fernández-Villa, C. Moreno-Iribas, V. Moreno, J. García-Pórez, G. Castaño-Vinyals, M. Pollán, M. Aubé, and M. Kogevinas, . 2019. 'Artificial light at night (alan), blue light spectrum exposure and colorectal cancer risk in Spain (MCC-Spain study).', *Environmental Epidemiology*, 3.

Garcia-Saenz, A., A. Sánchez-de Miguel, A. Espinosa, A. Valentin, N. Aragonés, J. Llorca, P. Amiano, V. Martín Sánchez, M. Guevara, R. Capelo, A. Tardón, Peiró-Perez, Iménez Moleón R.o., J.J., , A. Roca-Barceló, B. Pérez-Gómez, T. Dierssen-Sotos, T. Fernández-Villa, C. Moreno-Iribas, V. Moreno, J. García-Pérez, G. Castaño Vinyals, M. Pollán, M. Aubé, and M. Kogevinas. 2018. 'Evaluating the association between artificial light-at-night exposure and breast and prostate cancer risk in spain (MCC-spain study)', *Environmental Health Perspectives*, 126.

Gaston, K. J., T.W. Davies, J. Bennie, and J. Hopkins. 2012. 'Reducing the ecological consequences of night-time light pollution: options and developments', *Journal of Applied Ecology*, 49: 1256–66.

Gaston, K.J., S. Gaston, J. Bennie, and J. Hopkins. 2015. 'Benefits and costs of artificial nighttime lighting of the environment', *Environmental Review*, 23: 14–23.

Hölker, F., T. Moss, B. Griefahn, W. Kloas, C. Voigt, D. Henckel, A. Hänel, P. Kappeler, S. Völker, and A. Schwope. 2010. 'The dark side of light: a transdisciplinary research agenda for light pollution policy', *Ecology and Society*, 15.

Johnson, P.E., M.O. Smith, and J.B. Adams. 1985. 'Quantitative analysis of planetary reflectance spectra with principal components analysis', *Journal of Geophysical Research*, 90: C805-C10.

Johnson, P.E., M.O. Smith, S. Taylor-George, and J.B. Adams. 1983. 'A semiempirical method for analysis of the reflectance spectra fo binary mineral mixtures', *Journal of Geophysical Research*: 3557-61.

Kruse, F. A., and C. D. Elvidge. 2011. "Characterizing urban light sources using imaging spectrometry." In *2011 Joint Urban Remote Sensing Event, JURSE 2011 - Proceedings*, 149-52.

Levin, N., C. Kyba, Q. Zhang, A. Sánchez de Miguel, M. Román, X. Li, B. Portnov, A. Molthan, A. Jechow, S. Miller, Z. Wang, R. Shrestha, and C. Elvidge. 2020. 'Remote sensing of night lights: A review and an outlook for the future', *Remote Sensing of Environment*, 237.

Metcalf, Jeremy P. 2012. 'Detecting and characterizing nighttime lighting using multispectral and hyperspectral imaging', Thesis, Monterey, California. Naval Postgraduate School.

Moser, W. 2017. "Like It or Not, Chicago's About to Get a Lot Less Orange." In *Chicago Magazine*. Chicago IL.

Rich, C., and T. Longcore. 2013. *Ecological Consequences of Artificial Night Lighting* (Island Press).

Sánchez de Miguel, A., C.C. Kyba, M. Aubé, J. Zamorano, N. Cardiel, C. Tapia, J. Bennie, and K.J. Gaston. 2019. 'Colour remote sensing of the impact of artificial light at night (I): The potential of the International Space Station and other DSLR-based platforms', *Remote Sensing of Environment*, 224: 92-103.

Sánchez de Miguel, A., J. Zamorano, M. Aubé, J. Bennie, J. Gallego, F. Ocaña, D.R. Pettit, W. L. Stefanov, and K. J. Gaston. 2021. 'Colour remote sensing of the impact of artificial light at night (II): Calibration of DSLR-based images from the International Space Station', *Remote Sensing of Environment*, 264.

Settle, J. J., and N. A. Drake. 1993. 'Linear mixing and the estimation of ground cover proportions', *International Journal of Remote Sensing*, 14: 1159-77.

Singer, R. B., and T. B. McCord. 1979. "Mars: Large scale mixing of bright and dark surface materials and implications for analysis of spectral reflectance." In *10th Lunar and Planetary Science Conference*, 1835-48. American Geophysical Union.

Small, C. 2004. 'The Landsat ETM+ Spectral Mixing Space', *Remote Sensing of Environment*, 93: 1 –17.

———. 2018. "Multisource Imaging of Urban Growth and Infrastructure Using Landsat, Sentinel and SRTM." In *NASA Landsat-Sentinel Science Team Meeting*, edited by G. Gutman. Rockville MD: NASA.

———. 2019. 'Multisensor characterization of urban morphology and network structure', *Remote Sensing*, 11: 1-30.

Small, C., and C. Milesi. 2013. 'Multi-scale Standardized Spectral Mixture Models', *Remote Sensing of Environment*, 136: 442-54.

Small, C., A. Okujeni, S. van der Linden, and B. Waske. 2018. 'Remote Sensing of Urban Environments.' in S. Liang (ed.), *Comprehensive Remote Sensing* (Oxford: Elsevier).

Smith, M.O., S.L. Ustin, J.B. Adams, and A.R. Gillespie. 1990. 'Vegetation in deserts: I. A regional measure of abundance from multispectral images', *Remote Sensing of Environment*, 31: 1-26.

Sousa, D., and C. Small. 2017. 'Global cross calibration of Landsat spectral mixture models', *Remote Sensing of Environment* 192: 139–49.

———. 2019. 'Globally standardized MODIS spectral mixture models', *Remote Sensing Letters*, 10: 1018-27.

Stark, K. 2018. 'Chicago dials down LED street lamp intensity — and controversy'. https://energynews.us/2018/03/07/chicago-dials-down-led-street-lamp-intensity-and-controversy/.